\documentclass[aps,prl,reprint,superscriptaddress, longbibliography]{revtex4-1}
\usepackage{graphicx}
\usepackage{graphics}
\usepackage{epsfig}
\usepackage{epstopdf}
\usepackage{placeins}
\usepackage{lipsum}

\usepackage{array} 
\usepackage{multirow}

\usepackage{xcolor}
\usepackage[colorlinks,linkcolor=blue,citecolor=blue,urlcolor=blue] {hyperref}
\usepackage{amsmath}

\newcommand{\DDS}{$\mathrm{Dy_2ScN@C_{80}}$}

\newcommand{\TTS}{$\mathrm{Tb_2ScN@C_{80}}$}

\newcommand{\DDO}{$\mathrm{Dy_2O@C_{82}}$}
\newcommand{\Deff}{\Delta_\mathrm{eff}}

\begin{document}


\title{The sub-Kelvin hysteresis of the dilanthanide single molecule magnet Tb$_2$ScN@C$_{80}$}

\author{Aram Kostanyan}
\affiliation{Physik-Institut, Universit\"at Z\"urich, Winterthurerstrasse 190, CH-8057 Z\"urich, Switzerland}
\author{Rasmus Westerstr\"om}
\affiliation{Division of Synchrotron Radiation Research, Institute of Physics, University of Lund, SE-221 00 Lund, Sweden}
\author{David Kunhardt}
\affiliation{Leibniz Institute of Solid State and Materials Research, Dresden, D-01069 Dresden, Germany}
\author{Bernd B\"uchner}
\affiliation{Leibniz Institute of Solid State and Materials Research, Dresden, D-01069 Dresden, Germany}
\author{Alexey A. Popov}
\affiliation{Leibniz Institute of Solid State and Materials Research, Dresden, D-01069 Dresden, Germany}
\author{Thomas Greber}
\email{greber@physik.uzh.ch}
\affiliation{Physik-Institut, Universit\"at Z\"urich, Winterthurerstrasse 190, CH-8057 Z\"urich, Switzerland}


\date{\today}

\begin{abstract}
Magnetic hysteresis is a direct manifestation of non-equilibrium physics that has to be understood if a system shall be used for information storage and processing. 
The dilanthanide endofullerene \TTS\ is shown to be a single-molecule magnet with a remanence time in the order of 100~s at 400~mK. 
Three different temperature dependent relaxation barriers are discerned.
The lowest 1~K barrier is assigned to intermolecular interaction. The 10~K barrier to intramolecular exchange and dipolar coupling and the 50~K barrier to molecular vibrations as it was observed for \DDS . The four orders of magnitude difference in the prefactor between the Tb and the Dy compound in the decay process across the 10 K barrier is assigned to the electron number in the 4f shells that evidences lack of Kramers protection in Tb$^{3+}$.
The sub-Kelvin hysteresis follows changes in the magnetisation at adiabatic and non-adiabatic level crossings of the four possible Tb$_2$ ground state configurations as is inferred from a zero temperature hysteresis model.  

\end{abstract}

\maketitle

Single molecule magnets realize bistable spin configurations with lifetimes in the order of seconds or longer \cite{Gatteschi2006}. 
While their magnetisation may change via thermal fluctuations, there are as well temperature independent quantum flip mechanisms due to the tunneling of the magnetisation \cite{Thomas1996,Zhu2019}.
For the identification and use of quantum effects experiments have to be performed at sub-Kelvin temperatures, where temperature induced switching between different magnetic states is minimal. 

An important horizon in the research on single molecule magnets was reached by the discovery of hysteresis in double decker phthalocyanine complexes [Pc$_2$Ln], where one lanthanide ion is sandwiched by two organic moieties \cite{Ishikawa2003}.
With the terbium Pc$_2$Ln derivative it was shown later that the four nuclear spin levels of the terbium atom may be addressed and manipulated in a molecular break junction \cite{Wernsdorfer2012} and for dysprosium ions hysteresis up to 60 K was observed \cite{Goodwin2017}.
After having reached this fundamental limit of single ion magnetism the question on how two magnetic ions interact is an obvious continuation of exploration.
For the case of two holmium atoms on magnesium oxide at a separation distance of 1.2~nm the bistability of the individual Ho atoms appeared not to be influenced by the magnetic neighbourhood and they could be addressed as classical bits \cite{Natterer2017}.
In molecules containing two magnetic ions at closer distance, the exchange and dipolar interaction may lead to a preferred spin configuration \cite{rin11,Guo2011,Westerstrom2014,Liu2019,Yang2019}.
In order to examine the coupling between magnetic moments it is essential to have a stable, atomically precise environment of the spin system, and a geometry where the interaction is not negligible. 
Endohedral single molecule magnets meet these requirements since it is possible to place two lanthanides with distances below 0.4 nm in a C$_{80}$ cage \cite{Popov2013}. 
Before single single molecule magnet experiments are performed it is desirable to investigate ensembles of molecules. This allows an accurate determination of the magnetic lifetimes, and a faster screening for the "ideal" molecule. 
Here we report on the magnetisation of \TTS\ ensembles. 
The pseudospin model of Westerstr\"om et al. \cite{Westerstrom2014} may be successfully applied to the description of the electronic groundstate, though we find quantitative differences to \DDS . 
In \TTS\ the
zero field exchange protection as observed for \DDS\ with two Dy Kramers ions is four orders of magnitude weaker if the zero field lifetime is taken as a measure for this protection.
This is also reflected in the hysteresis, i.e. memory of the magnetisation history. 
Hysteresis occurs whenever the magnetisation in a field scan does not follow the adiabatic ground-state.
The hysteresis depends on the lifetime of a given magnetisation for a given applied field.
This explains on why a quantitative prediction of the hysteresis curve for single molecule magnets is more involved than the description of the magnetisation in thermal equilibrium.
Steps and kinks in hystereses are often associated to level crossings \cite{Thomas1996}.
Since the level crossings depend on the applied field vector and the anisotropy axes, single crystal and single molecule experiments at lowest temperatures display the sharpest steps.
Magnetisation hysteresis of rare earth ions substituted in LiYF$_4$ and in molecule-crystals has been investigated at sub-Kelvin temperatures before \cite{Giraud2001,Blagg2013}. 
In the present letter we report sharp steps in the hysteresis for an anisotropic \TTS\ powder sample. These steps are related to the crossings of quantum levels with different spin configuration that occur in a narrow external magnetic field window. The simplicity of the Tb$_2$ spin configuration in the present molecule allows the modelling of the zero temperature hysteresis with non-adiabatic ground state level crossings and a quantitative comparison to the 
observed hysteresis curve. 

\begin{figure}[b]
    \centering
    \includegraphics[width=\columnwidth]{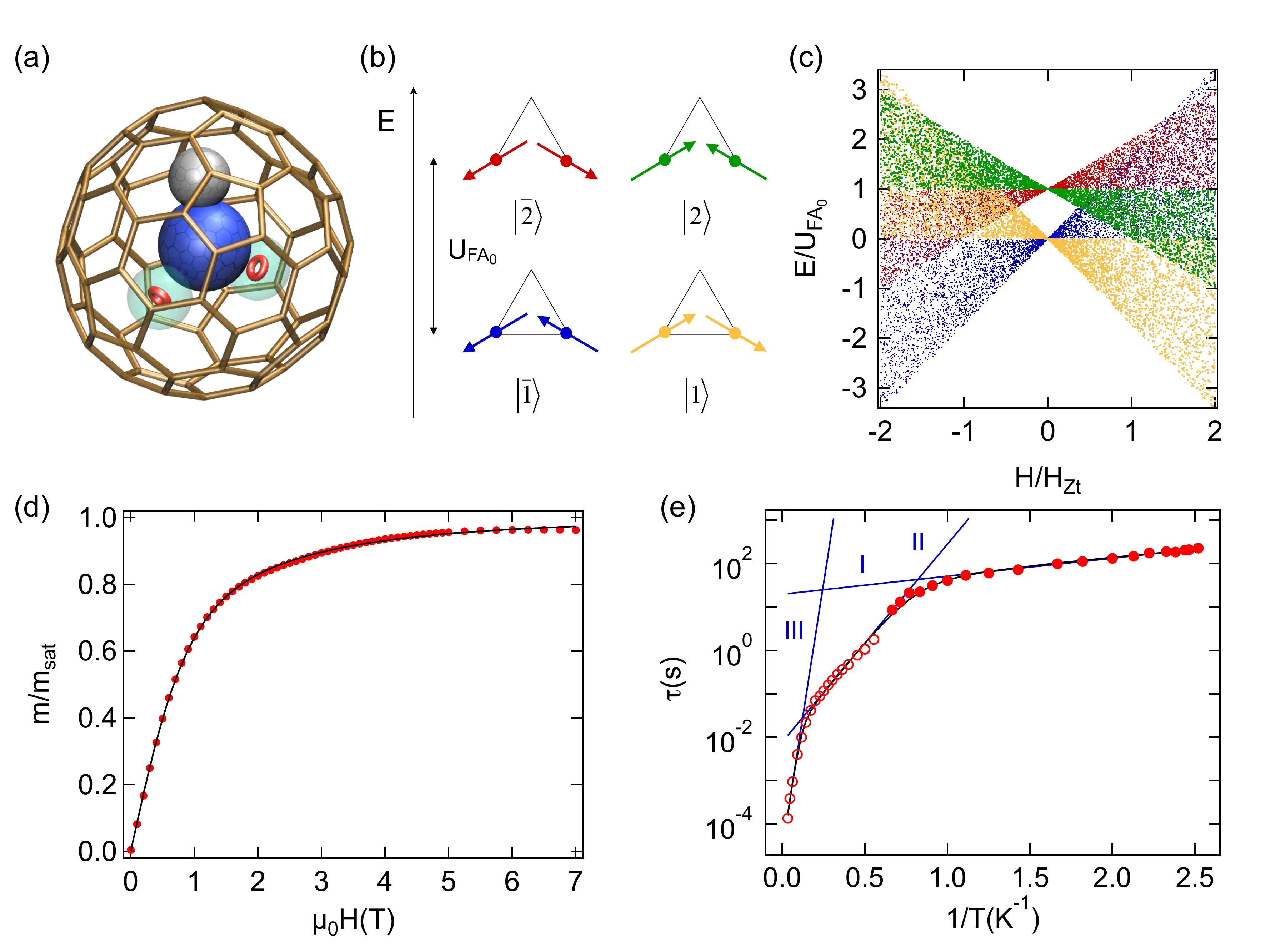}
    \caption{(a) Model of the dilanthanide single molecule magnet \TTS . The sizes of the endohedral ions is mimicked with their ion radii. In the center the of the Tb$^{3+}$ ions (turquois) the 4f ($J_z$ = 6) orbitals that constitute the paramagnetism are depicted in red \cite{Sievers1982,Greber2019}. (b) Magnetic zero field ground states: The magnetic moments $\mu$ of the two Tb ions sit on an equilateral triangle and point into the center or away from it. Two doublets form, where $ |{\bar{1}}\rangle$ (blue) and $|1\rangle$ (yellow) have ferromagnetic coupling, while  $|{\bar{2}}\rangle$ (red) and $|2\rangle$ (green) are antiferromagnetically coupled. In zero field the energy difference between the two doublets is U$_{{\rm{FA}}_0}$.  
 (c) Energies for the four ground states of 10~000 molecules oriented randomly with respect to the axis of the applied H-field. The color coding is adopted from (a). No level crossings between the two doublets are expected for $|{\rm{H}}|< {\mathrm{U}} _{{\rm{FA}}_0}/2\mu_0 \mu$ \cite{supplementals}.
(d) Magnetisation curve at 6~K and fit of the pseudospin model (black line) after diamagnetic background subtraction resulting in $\mu=8.8~\mu_B$ and U$_{{\rm{FA}}_0}/k_B$=9.4~K.
(e) Arrhenius plot of the magnetisation lifetimes. Full symbols DC measurements with the $^3$He cryostat, open symbols AC susceptibility measurements. The black curve is the sum of three Arrhenius processes, I, II and III (the slopes of the blue lines represent the individual process barriers). 
}
    \label{F1}
\end{figure}

\TTS\ (see Figure \ref{F1}(a)) endofullerenes were produced by arc-discharge synthesis using the corresponding metals \cite{Zhang2015}.
For the magnetization measurements, the molecule - toluene solution was drop-cast into a poly propylene sample holder resulting in a visible black powder residue. 
From the saturation magnetization of 
$2.98 \times 10^{17} \mu_B$ and an average molecular moment of 9~$\mu_B$ an ensemble of \(3.3 \times 10^{16}\) 
molecules or 73.5~$\mu$g 
is inferred.
The magnetization was measured in a Quantum Design MPMS3 Vibrating Sample Magnetometer (VSM) with a $^{3}$He cryostat.
AC magnetisation measurements were performed in the temperature range between 1.8~K and 30~K at zero \mbox{($<0.5$~mT)} DC magnetic field with a driving AC field amplitude of 1~mT up to 10~Hz and 0.25~mT between 10~Hz and 1~kHz.


Figure \ref{F1}~(b) shows the model of the ground state of \TTS\ in zero external magnetic field. 
The two paramagnetic Tb$^{3+}$ ions in the Tb$_2$ScN endohedral unit constitute the molecular magnetism.
The degeneracy of the eight 4f electrons in the $^{7}$F$_{6}$ Hund manifold is lifted in the ligand field that is dominated by the central N$^{3-}$ ion.
As for Dy \cite{Westerstrom2014} and Ho \cite{Dreiser2014} the maximum projections of $J$ assume the groundstates, which are in the case of terbium the $J_z=\pm6$ levels with a nominal magnetic moment of $\mu=\pm 9~ \mu_B$ along the Tb--N axes.
The magnetic anisotropy is high and other $J_z$ states may be neglected since they have energies that are much higher than the thermal energies in the present experiments \cite{Cimposeu2014}. 
Below 50~K, the molecule orientation is frozen, and the ground states of the individual molecules are determined by the orientation of the external magnetic field with respect to that of the magnetic moments on the two Tb atoms \cite{Fu2011, Kostanyan2017}.
For a given molecule this yields 2$^2$ possible ground state configurations that split into two time reversal symmetric doublets spanning the Hilbert space \cite{Westerstrom2014}. 
The states $|1\rangle, |{\bar{1}}\rangle$ are ferromagnetically and $|2\rangle,|{\bar{2}}\rangle$ are antiferromagnetically coupled and the energy difference U$_{{\rm{FA}}_0}$ is reflected in the magnetisation curves.
In \TTS\ $|1\rangle$ and $|{\bar{1}}\rangle$ have lower energy in zero field than $|2\rangle$ and $|{\bar{2}}\rangle$ as it was found for \DDS\ \cite{Westerstrom2014}, while e.g. in \DDO\ antiferromagnetic coupling is favored \cite{Yang2019}.
For Tb ions sitting on two vertices of an equilateral triangle the total magnetic moments of $|1\rangle, |{\bar{1}}\rangle$ and $|2\rangle, |{\bar{2}}\rangle$ are orthogonal and $\pm \sqrt{3} \mu$ for the ferromagnetic  and $\pm \mu$ for the antiferromagnetic doublet.
The energy difference between the two doublets has exchange and dipolar components and
in an external magnetic field the degeneracies of the doublets are lifted by the corresponding Zeeman splitting.
The magnetism is non-collinear i.e. the magnetic moments are not aligned to the external field but to the molecular coordinates that determine the anisotropy axes.
In zero field cooled powder samples there is no preferential molecular orientation and the distribution of the Tb-N axes is isotropic \cite{Kostanyan2017}.
In Figure~\ref{F1}(c) the energies of an ensemble of isotropically distributed molecules in different external magnetic fields are displayed.
The energy and field scales in Figure~\ref{F1}(c) are U$_{{\rm{FA}}_0}$ and the Zeeman threshold field above which antiferro states with an according orientation in the field may get the lowest energy H$_{\rm{Zt}}\equiv \rm{U}_{{\rm{FA}}_0}/\mu_0\mu$.

In Figure~\ref{F1}(d) the equilibrium magnetisation for \TTS\ at 6~K with a corresponding fit of the pseudospin model \cite{Westerstrom2014} is shown. The fit yields a Tb magnetic moment $\mu$ of 8.8$\pm0.4~\mu_B$, and an exchange and dipolar barrier U$_{{\rm{FA}}_0}/k_B$ of 9.4$\pm$1.5~K. 
\begin{figure*}[t]
\centering
\includegraphics[width=15cm]{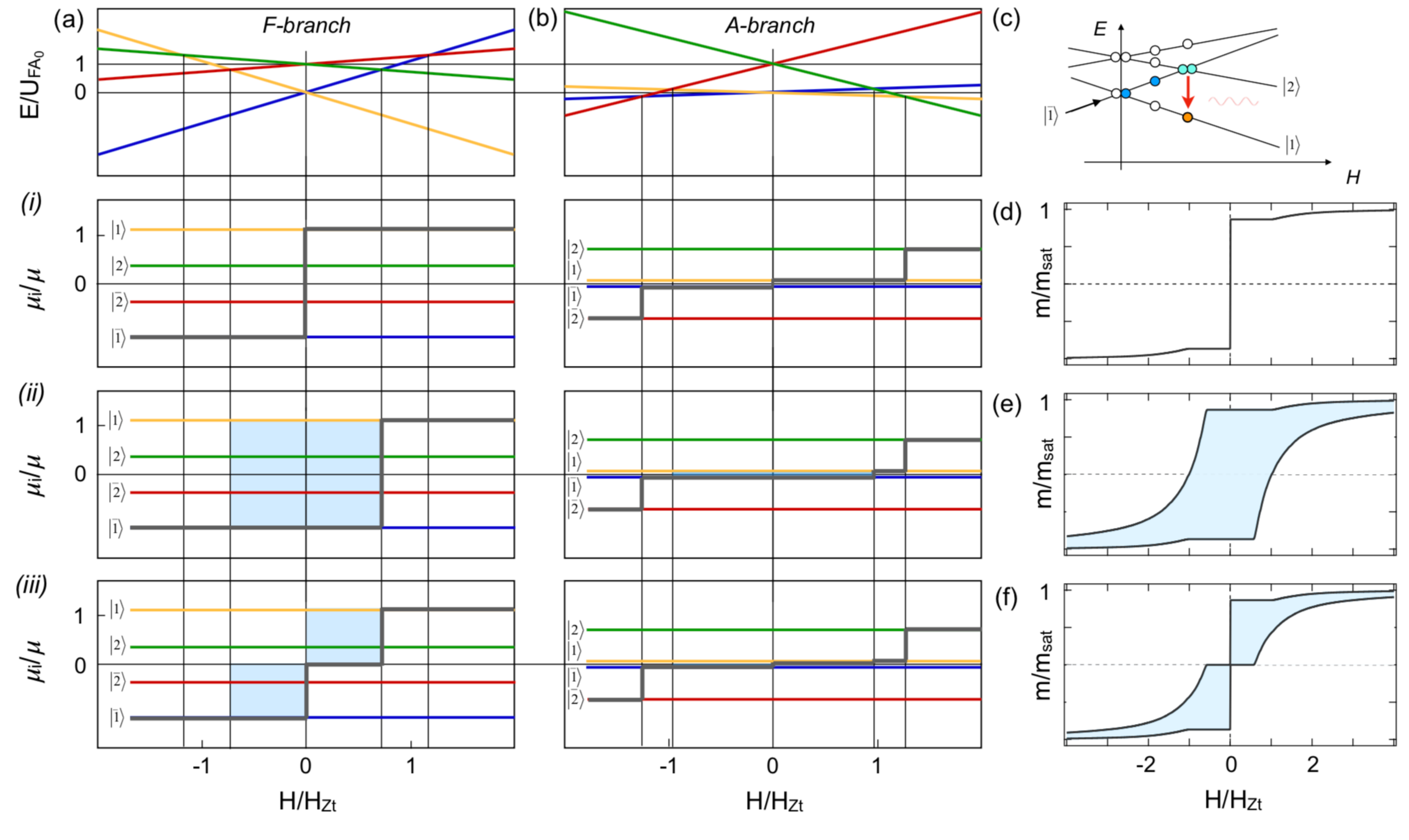}
\caption{The 6 different level crossings within the four ground states $|1\rangle, |{\bar{1}}\rangle, |2\rangle, |{\bar{2}}\rangle$ (yellow, blue, green, red) in a Zeeman E~vs.~H diagram and possible zero temperature magnetisation curves. a) Example of an F-branch orientation with ferromagnetic ground states. b) Example of an A-branch orientation with antiferromagnetic ground states in large fields. The bottom panels show the magnetisation as expected for different scenarios {\it{(i), (ii)}} and {\it{(iii)}} (for details see text). c) Crossing that reassumes non-adiabatically the ground state with energy dissipation for a field scan from negative to positive external fields. Upon crossing of the $|{\bar{1}}\rangle$ and the $|2\rangle$ state one spin may flip without cost of energy, and the second spin may flip under the release of energy into the $|1\rangle$ state. (d-f) Magnetisation curves of the sum of randomly oriented molecules for the three scenarios. While the adiabatic scenario  {\it{(i)}} displays no hysteresis, the two with non-adiabatic jumps in the magnetisation do.}

\label{F2}
\end{figure*}
These parameters determine the Zeeman threshold field $\mu_0\rm{H_{\rm{Zt}}}$=1.6$\pm$0.3 T.
It is known that U$_{{\rm{FA}}_0}$ is also reflected in non-equilibrium data as it is the decay time to reach equilibrium \cite{Westerstrom2014}.
Figure~\ref{F1}(e) displays the zero field magnetisation lifetimes of  \TTS\  in an Arrhenius plot in the temperature range between 0.4 and 30~K \cite{supplementals}.
From the fit 
three different decay processes with barriers $\Delta_{{\rm{eff}}}^i$ and prefactors $\tau_{0,i}$ as listed in Table \ref{T1} are inferred.
\begin{table}[ht]
\centering
\begin{tabular}{ l |c |r }
process & $\tau_{0}$~(s)& $\Deff/k_B$~(K) \\
\hline
I & $(2.0 \pm 0.3)\times 10^{1}$~~~& $1 \pm 0.1$ \\
II & $(7.7 \pm 0.1)\times 10^{-3} $    & $10.5 \pm 0.3$ \\
III & $(2.6 \pm 0.5)\times 10^{-5} $ & $56.4 \pm 3.0$ \\
\end{tabular}
\caption{Fit parameters of three Arrhenius barriers for the temperature dependence of the zero field magnetization relaxation times $\tau$ of \TTS\ in Figure~\ref{F1}(e). {\mbox{$\tau^{-1}=\sum_{i=I}^{III} \tau_{0,i}^{-1}\exp{(-\Delta^i_{\rm{eff}}/k_BT)}$.}} }
\label{T1}
\end{table}
Process II is identified as the decay that is mediated via the excitation across U$_{{\rm{FA}}_0}$, and $\Delta_{\rm{eff}}^{II}$ of 10.5~K is comparable to that of \DDS\ (8.5 K) \cite {Westerstrom2014}. On the other hand, the prefactor of \TTS\ is four orders of magnitude smaller. 
This must be related to the even symmetry of the Tb~4f$^8$ and the odd Kramers symmetry of Dy~4f$^9$.
The 1~K barrier may not be explained within the ground state picture in Figure \ref{F1}(b). 
It rather points to dipolar intermolecular interactions as they were e.g. proposed to explain a 1.2~K transition in a Fe$_{19}$ nanodisk system \cite{Pratt2014}. 
For the present case of close packed \TTS\ with randomly oriented endohedral units we get from Monte Caro simulations dipolar interaction energy distributions with a full width at half maximum of 0.9 $k_B$~K, which is very close to the observed barrier $\Deff^{I}$ of 1~K.
The barrier of the fastest process III is similar to a barrier in \DDS\ \cite{Westerstrom2014} and likely involves molecular vibration assisted transitions.

Now we develop a theory for hystereses without thermal relaxation of the magnetisation.
Figure~\ref{F2} shows the Zeeman energies for two molecular orientations relative to the external field of the $|1\rangle, |{\bar{1}}\rangle$ and the $|2\rangle, |{\bar{2}}\rangle$ doublets and their corresponding contribution to the magnetisation. 
The majority group with ferromagnetic ground states at all fields $|1\rangle$ or $ |{\bar{1}}\rangle$ is called F-branch and depicted in (a). The minority group with antiferromagnetic ground states in fields above H$_{\rm{Zt}}$ or below -H$_{\rm{Zt}}$, $|2\rangle$ or $|{\bar{2}}\rangle$, is called A-branch and depicted in (b).
At level crossings the magnetisation of the given molecules may flip between two values without cost or release of energy. 
This is visualized in the bottom panels, where the different effective magnetic moments $\mu_{i}=d{\rm{E}}_i/d\mu_0\rm{H}$ and their occupancy are displayed. 
Both branches have six level crossings, though the topology, or crossing sequence is different for the F- and the A-branch: 
We distinguish two zero field crossings: $|{\bar{1}}\rangle \leftrightarrow |1\rangle$ and $|{\bar{2}}\rangle \leftrightarrow |2\rangle$, which involve the simultaneous flip of the two spins that constitute the state, and four non-zero field crossings: $|{\bar{2}}\rangle \leftrightarrow |{\bar{1}}\rangle$, $|{\bar{2}}\rangle\leftrightarrow |1\rangle$, $|{\bar{1}}\rangle \leftrightarrow |2\rangle$ and  $|1\rangle \leftrightarrow |2\rangle$ that involve one spin flip, only.
For scenarios where the change in magnetisation occurs at level crossings only, we construct the zero temperature magnetisation curve of the ensemble from the sum of the projected (effective) magnetic moments of all molecular orientations. 
If the magnetisation in a field-scan would follow the lowest energy, no hysteresis were expected because the Zeeman energy diagram is time reversal symmetric.
In order to observe hysteresis we rely on crossings where the system prevails in its magnetisation state and leaves the lowest energy curve, and where it jumps in subsequent crossings non-adiabatically to the lowest energy. 
In Figure~\ref{F2} the magnetisation is shown for three scenarios and the F- and the A-branch, where the field-scan starts at saturation in large negative fields i.e. $|{\bar{1}}\rangle$ or $|{\bar{2}}\rangle$: {\it{(i)}} lowest energy, {\it{(ii)}} crossing at zero field and non-adiabatic relaxation upon the single flip crossing $|{\bar{1}}\rangle \rightarrow |2\rangle \searrow |1\rangle$, and {\it{(iii)}}  equilibration at zero field into 50\% $|1\rangle$ and 50\% $|{\bar{1}}\rangle$ and non-adiabatic relaxation at the single flip crossing $|{\bar{1}}\rangle \rightarrow |2\rangle \searrow |1\rangle$.
Figure \ref{F2}(c) shows the crossing and non-adiabatic relaxation $|{\bar{1}}\rangle \rightarrow |2\rangle \searrow |1\rangle$ for the relaxation of excited states.
The magnetisation in the adiabatic scenario {\it{(i)}} displays no hysteresis (see Figure \ref{F2}(d)), but the deviation from the step function clearly discerns the influence of the A-branch and the "gap" between $\pm \rm{H_{\rm{Zt}}}$, where no adiabatic single flip crossings occur. 
For the two scenarios with non-adiabatic crossings Figure \ref{F2}(e,f) the state distribution depends on the field scan direction, and correspondingly hysteresis is expected.
Importantly, we see that $|{\bar{1}}\rangle \rightarrow |2\rangle \searrow |1\rangle$ crossings of excited states may occur in the ground state crossing gap for H$>\rm{H_{\rm{Zt}}}/2$ and that these transitions sharply peak above $\rm{H_{\rm{Zt}}}/2$ \cite{supplementals}. This theory provides an upper limit of the deviation of the magnetisation from the equilibrium and is therefore a benchmark for the characterisation of the hysteresis of dilanthanide single molecule magnets. 

\begin{figure}[!ht]
    \centering
    \includegraphics[width=9 cm]{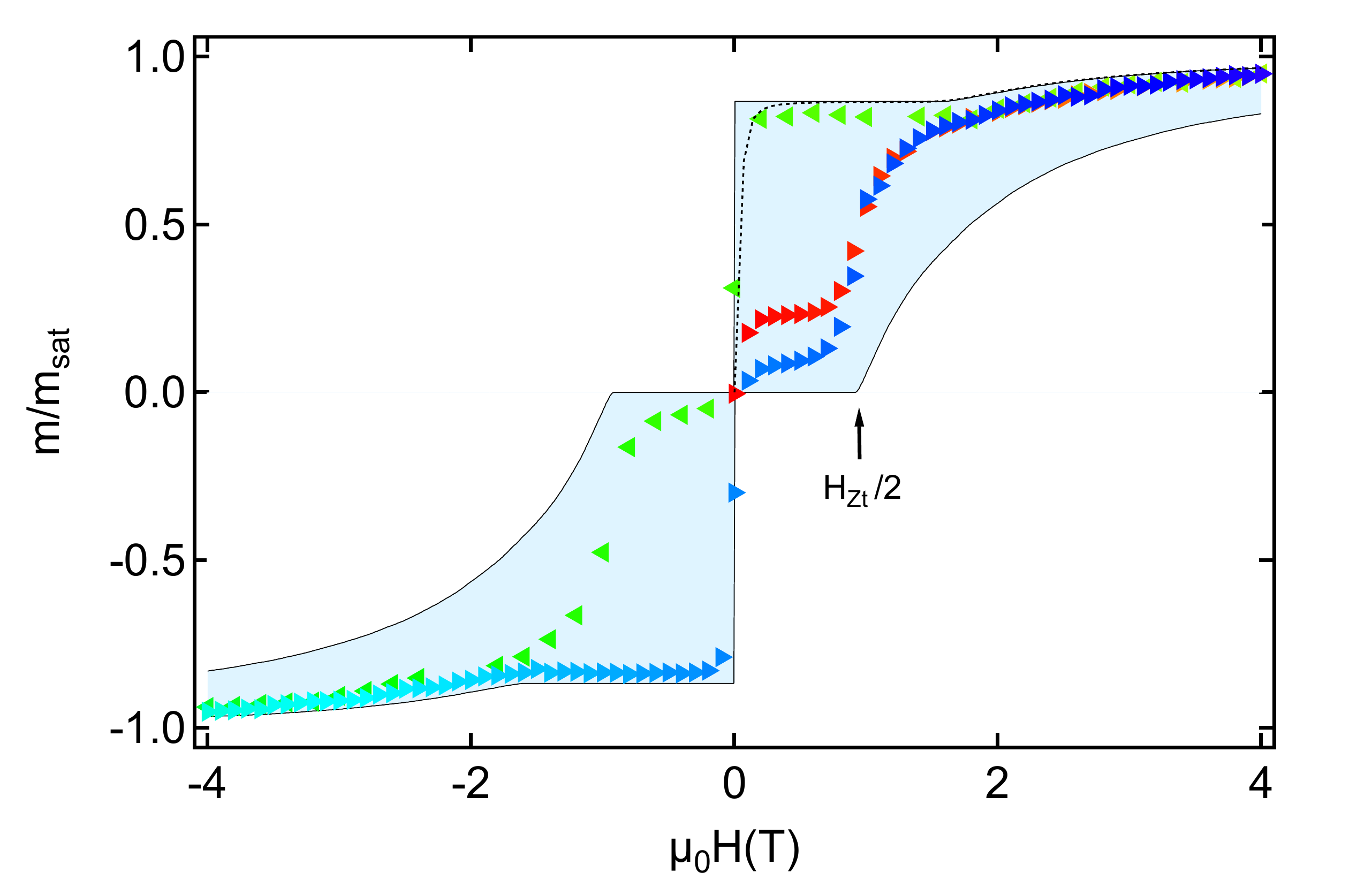}
    \vspace{-0.5cm}
    \caption{Magnetisation loop of \TTS\ recorded at 390~mK with a field scan rate of $\pm$3.3 mT/s. The red green blue color code of the experimental data represents the time during the field scan that starts with the virgin curve at H=0 (red). The triangles indicate the positive ($\triangleright$) and negative ($\triangleleft$) field scan rate. The light blue area represents the zero temperature hysteresis of scenario {\it{(iii)}}. H$_{\rm{Zt}}$/2 is the magnetic field at which onset of non-adiabatic decay of magnetisation is expected. The dotted line is the theoretical equilibrium magnetisation curve at 390~mK.}
    \label{F3}
\end{figure}

The three scenarios in Figure \ref{F2} may be compared to sub-Kelvin magnetisation data.
Figure \ref{F3} shows the magnetisation curve for \TTS\ at 390~mK with a field scan rate of 3.3 ~mT/s. Starting at zero field the magnetisation jumps within 60~s to 20\% of the saturation magnetisation where it remains constant  before it continues to rise at 0.75~T external field.
This can be understood within the ground state picture of Figure \ref{F1}(b). Near zero field fluctuations between $|1\rangle$ and $ |{\bar{1}}\rangle$ states prevail 
and the rise of the external field increases the magnetisation. 
Between $\mu_0$H of 0.2 and 0.7~T the fluctuations are suppressed due to Zeeman energies exceeding $k_BT$ and the magnetisation appears to be frozen. At  \mbox{$\mu_0 \rm{H}$=0.75 T} the magnetisation rises again. 
This field corresponds to $\mu_0\rm{H_{\rm{Zt}}}/2$ and is a confirmation that $|{\bar{1}}\rangle \rightarrow |2\rangle \searrow |1\rangle$ transitions drive the increase of magnetisation towards saturation. If the field scan direction is inverted the system remains in a high magnetisation state down to zero field where equilibration is most effective, and where the magnetisation decays.
The observed hysteresis compares best with scenario {\it{(iii)}}, where we observe a kink in the lower branch of the magnetisation curve at half the adiabatic threshold field H$_{\rm{Zt}}$. Still, the measured hysteresis indicates less magnetisation hysteresis than would be expected if level crossings alone would cause changes in the magnetisation of the sample. 
This indicates that at 400 mK other processes still contribute to the decay of the magnetisation toward the thermal equilibrium \cite{Zhu2019}.

In conclusion \TTS\ is shown to be a single molecule magnet with a ground state configuration that causes a sub-Kelvin hysteresis with kinks, which coincide with adiabatic zero field crossings and non-adiabatic non-zero field crossings at a characteristic external magnetic field. The findings can be translated to any single molecule magnet with two spin centers.

Financial support from the Swiss National Science Foundation (SNF project 200021 129861, 147143, and PZ00P2-142474), the European UnionÕs Horizon 2020 research and innovation program, European Research Council (grant agreement No 648295 to A.A.P.), the Deutsche Forschungsgemeinschaft (DFG project PO 1602/4-1 and 1602/5-1) and the Swedish Research Council (Grant No. 2015-00455) are acknowleged. We thank \mbox{Ari P. Seitsonen} for the artwork in Figure~\ref{F1}(a).

\bibliography{Tb2Sc}
\end{document}